\documentclass[reprint,aip,jcp]{revtex4-1}

\usepackage{amsmath}
\usepackage{graphicx}
\usepackage{color}

\newcommand{\Ket}[1]{\vert#1\rangle}
\newcommand{\Bra}[1]{\langle#1\vert}
\newcommand{\T}[1]{\mathrm{#1}}
\newcommand{\V}[1]{\boldsymbol{#1}}
\newcommand{\Op}[1]{\hat{#1}}
\newcommand{\BS}[1]{({#1})}

\begin{document}

\title{Vibronic exciton theory of singlet fission.~II.~Two-dimensional spectroscopic detection of the correlated triplet pair state}

\author{Roel Tempelaar}
\email{r.tempelaar@gmail.com}
\author{David R.~Reichman}
\email{drr2103@columbia.edu}
\affiliation{Department of Chemistry, Columbia University, 3000 Broadway, New York, New York 10027, USA}

\begin{abstract}
Singlet fission, the molecular process through which photons are effectively converted into pairs of lower energy triplet excitons, holds promise as a means of boosting photovoltaic device efficiencies. In the preceding article of this series, we formulated a vibronic theory of singlet fission, inspired by previous experimental and theoretical studies suggesting that vibronic coupling plays an important role in fission dynamics. Here, we extend our model in order to simulate two-dimensional electronic spectra, through which the theory is further validated based on a comparison to recent measurements on pentacene crystals. Moreover, by means of such spectral simulations, we provide new insights into the nature of the correlated triplet pair state, the first product intermediate in the fission process. In particular, we address a controversy in the literature regarding the identification, energies, and transition dipole moments of its optical transitions towards higher-lying triplet states.
\end{abstract}

\maketitle

\section{Introduction}

The Shockley-Queisser efficiency limit for single-junction solar cells\cite{Shockley_61a} can be circumvented\cite{Hanna_06a} by a molecular process called singlet fission in which an optically excited singlet state is converted into a pair of triplet excitons, which are transiently correlated in a spin-zero state. The prospect of enhancing photovoltaic efficiencies through such an exciton multiplication mechanism has stimulated intense interest in singlet fission.\cite{Smith_10a, Smith_13a} Recent time-resolved spectroscopic studies have provided indications that vibronic coupling plays an integral role in the excited state dynamics of singlet fission materials.\cite{Musser_15a, Bakulin_16a, Monahan_16a} However, a detailed understanding of the mechanisms through which singlet fission occurs remains lacking, and calls for a theoretical study in which a microscopic set of electronic degrees of freedom interacting with vibrational modes are treated non-perturbatively. In the preceding article of this series, hereafter referred to as I, we formulated a vibronic exciton model tailored to meet this demand. Using crystalline pentacene as a prototypical fission material we characterized the excited states contributing to linear absorption, paying special attention to the energy and spatial configuration of the correlated triplet pair state (TT$_1$), which acts as the first conversion intermediate in singlet fission.

The experimental detection of the fission product states is hindered by the fact that they are essentially optically dark with respect to the ground state, a feature that results from their multi-exciton character and the triplet nature of their constituents. On the picosecond time scale, singlet fission can be observed through recombination into singlet excitons by means of delayed fluorescence.\cite{Burdett_10a, Burdett_11a, Burdett_12a, Roberts_12a} Transient absorption\cite{Rao_10a, Burdett_10a, Wilson_11a, Rao_11a, Burdett_11a, Ramanan_12a, Roberts_12a, Musser_15a, Pensack_16a} and two-photon photoemission spectroscopy,\cite{Chan_11a, Chan_12a, Monahan_16a} on the other hand, allow for a sufficient time resolution to resolve the sub-ps fission dynamics in materials such as crystalline pentacene. Accordingly, it is assumed that the product states may be detected through allowed transitions towards other (auxiliary) excited states, typically higher-lying triplets or ionized states. However, the disadvantage of such approaches is that the optical signal simultaneously bears the imprints of both the fission product and the auxiliary states. For example, optical transitions between TT$_1$ and higher-lying triplet states inform on the relative energy difference of the involved states, but do not reveal their energetic separation from the ground state or the singlet excitation through which the fission process is initiated. This magnifies the issue that such triplet signals are commonly mapped onto the bright response from the singlet state, rendering the overall spectra congested with features. Perhaps for this reason, some of the identified triplet transitions in bulk pentacene\cite{Marciniak_07a, Marciniak_09a, Thorsmolle_09a, Thorsmolle_09b, Rao_10a, Rao_11a, Wilson_13a, Bakulin_16a} seem inconsistent with measurements\cite{Hellner_72a, Ashpole_74a} and calculations on non-interacting pentacene molecules,\cite{Pabst_08a} discrepancies that remain to be clarified.\cite{Smith_10a, Smith_13a}

Most of the aforementioned complications can be circumvented through two-dimensional electronic spectroscopy (2DES),\cite{Hybl_98a} which by projecting the optical response to an extra dimension allows one to resolve otherwise overlapping spectral signals.\cite{Mukamel_1, Jonas_03a, Cho_08a} 2DES has served as an important experimental technique for the study of sub-ps quantum dynamics in photosynthesis\cite{Brixner_05a, Zigmantas_06a, Engel_07a, Collini_10a, Fuller_14a, Romero_14a, Dostal_16a} as well as synthetic materials\cite{Collini_09a, Halpin_14a} for over a decade, but only recently found its first application in the context of singlet fission in a study on polycrystalline pentacene by Bakulin \textit{et al.}\cite{Bakulin_16a} One of the key results of that study was the direct detection of the optical transition between TT$_1$ and the ground state. Although this transition is extremely weak, this observation was accomplished by creating a coherent vibrational wavepacket in the pair state potential, yielding spectral beats that were enhanced through intensity borrowing. This experimental result helped clarify the energetic separation of TT$_1$ from the ground and singlet excited states, and was used in I to parametrize properties of the triplet excitons in the proposed model for pentacene crystals. By further analyzing the 2DES measurements through our model, additional crucial information about the excited states involved in singlet fission is readily obtainable.

In the present article, we extend our theoretical model in order to simulate 2DES of fission materials. This allows us to subject the model to additional tests by comparing simulations to the 2DES measurements on polycrystalline pentacene. Perhaps more importantly, however, it allows us to further investigate the microscopic nature of TT$_1$ through its contribution to time-resolved spectroscopy. In particular, by reproducing 2DES features associated with this state, we are able to shed light on its optical transitions towards higher-lying triplet states. In doing so, we address questions regarding the identification of these transitions, while at the same time describing important molecular properties that impact estimates of the associated signal strengths.

This paper is organized as follows. After a few remarks on the observation of singlet fission through time-resolved spectroscopy, we review in Sec.~\ref{Sec_Theory} the main aspects of the theory introduced in I. Subsequently, the necessary extensions of the vibronic basis set and Hamiltonian are discussed, followed by the theoretical framework employed to simulate 2DES. The section closes with a summary of the associated parameters for pentacene crystals. In Sec.~\ref{Sec_Results}, we present results obtained for crystalline pentacene through the model. First, static 2D spectra are compared to experimental measurements, as well as oscillatory spectra which serve to detect TT$_1$. What follows is an evaluation of the optical transitions between triplet states, through which TT$_1$ becomes visible in time-resolved spectra, and a discussion of several material properties which affect the strengths of these transitions. We conclude in Sec.~\ref{Sec_Conclusions}.

\section{Theory}\label{Sec_Theory}

\begin{figure}
\includegraphics{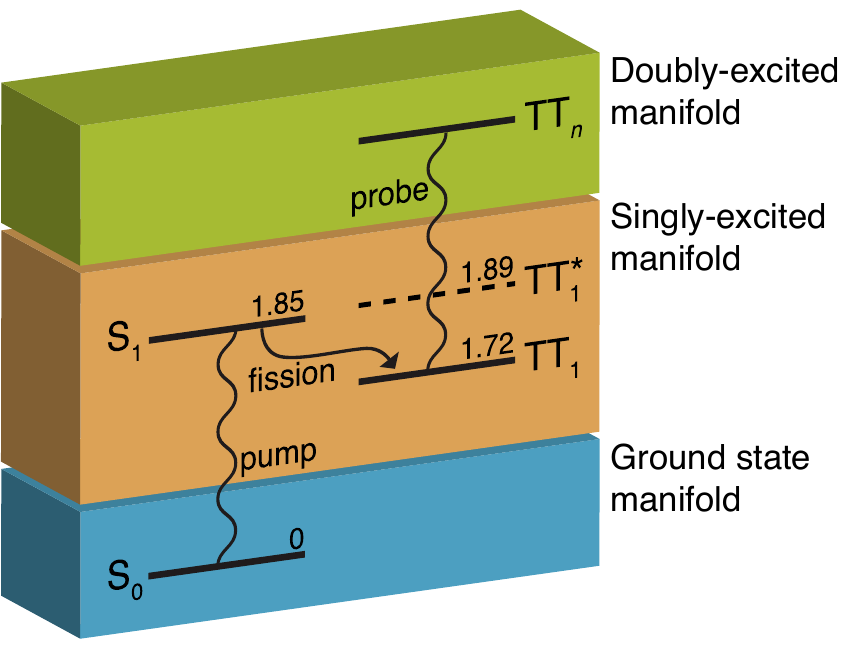}
\caption{Minimal representation of the fission of a singlet excitation into a correlated triplet pair state as observed through time-resolved spectroscopy. Shown are the adiabatic energy levels of the involved excited states: ground state (S$_0$), singlet excited state (S$_1$), and the low- and high-energy triplet pair states (TT$_1$ and TT$_n$). Two optical excitation events are indicated with wiggling lines. The first event (``pump'') brings the molecular system from the ground state to S$_1$. This triggers the fission process (arrow). The second event (``probe'') couples the product TT$_1$ state to a triplet pair state that is higher in energy. Together, the states span three different manifolds, between which transitions are only significant when induced by an optical pulse. The labels ``singly-excited'' and ``doubly-excited'' refer to the number of optical excitation events required to populate the manifolds starting from the ground state, and are not to be confused with the number of involved electron-hole pairs (excitons). For example, TT$_1$ is sometimes referred to as a double excitation, being composed of two excitons, but nevertheless resides in the singly-excited manifold as defined here. Numbers indicate energies (in eV) obtained for pentacene crystals in our preceding article. Also depicted is the vibrationally dressed triplet pair state (TT$_1^*$), found to be quasi-resonant with S$_1$ in bulk pentacene (dash) in Ref.~\citenum{Bakulin_16a} and in the preceding article.}
\label{Fig_FissionScheme}
\end{figure}

In order to streamline the discussion, we begin by showing in Fig.~\ref{Fig_FissionScheme} a schematic representation of how sub-ps fission dynamics is typically observed within time-resolved spectroscopy. This illustration is minimal in the sense that it represents the many excited states found in fission materials by a maximally reduced set: a singlet ground state (S$_0$), a singlet excited state (S$_1$), a correlated triplet pair (TT$_1$), and an analogues state that is higher in energy (TT$_n$). These are assumed to be adiabatic states, and as such belong to the quantum basis that is spectroscopically detectable. In a similarly simplified fashion, time-resolved spectroscopic experiments can be regarded as consisting of two optical excitation events separated in time by some varying interval. The first event (commonly referred to as ``pump'') excites the molecular system from its initial state S$_0$ to the excited state S$_1$. Subsequently, some non-adiabatic mechanism induces a transition from S$_1$ to TT$_1$, after which a second excitation event (the ``probe'') brings the system to the higher-lying pair state, TT$_n$. The observed signal associated with this second transition is used as a time-dependent probe of the fission process.\cite{Rao_10a, Burdett_10a, Wilson_11a, Rao_11a, Burdett_11a, Ramanan_12a, Roberts_12a, Musser_15a, Pensack_16a} We note that experiments using two-photon photoemission spectroscopy involved ionized states instead of TT$_n$.\cite{Chan_11a, Chan_12a, Monahan_16a} We will exclude such states from our discussion, and as such limit ourselves to (third-order) purely optical techniques such as 2DES and transient absorption.

The states depicted in Fig.~\ref{Fig_FissionScheme} span three manifolds, referred to as ground state, singly-excited, and doubly-excited (the last two names refer to the number of optical excitation events required to populate the manifolds starting from the ground state, not to the number of electron-hole pairs). Electronic couplings between states from different manifolds are usually negligible, and hence transitions between them are only significant when induced by an electromagnetic field. Nevertheless, rather than the minimal set of states depicted in Fig.~\ref{Fig_FissionScheme}, each manifold in reality consists of a large number of coupled states. In I, we characterized the excited states constituting the singly-excited manifold of crystalline pentacene, while making a connection to the simplified representation of singlet fission as shown in Fig.~\ref{Fig_FissionScheme}. This yielded the adiabatic energies reported in this figure, where it should be noted that TT$_1$ was found to consist of a handful of nearly-degenerate states, in contrast to the single bright state that represents S$_1$. Also depicted is the vibrationally dressed analogue of TT$_1$ (TT$_1^*$), which appears prominently in our results, and which was emphasized earlier by Bakulin \textit{et al.}\cite{Bakulin_16a} The underlying nearly-degenerate states are found to be quasi-resonant with S$_1$, and have been predicted to be of great importance to the fission dynamics in crystalline pentacene.\cite{Bakulin_16a}

Finally, we note that the non-adiabatic dynamics referred to in Fig.~\ref{Fig_FissionScheme} will be a topic of a future study, and is not included in the modeling presented here. Although the contribution of dynamical fluctuations are not present in the calculated spectra, our approach allows one to reproduce the detection of the correlated triplet pair state through a coherently prepared vibrational wavepacket,\cite{Bakulin_16a} which is the primary focus of the present study.

\subsection{Basis set and Hamiltonian}

In I, we presented in detail a diabatic vibronic basis set tailored to provide a description of the singly-excited manifold. The electronic subspace of this basis comprises a microscopic set of singlet states (s$_1$), triplet states (t$_1$), cationic states (c), and anionic states (a), where triplets always come in pairs and cations and anions always appear together, such that all basis states have zero spin and are charge neutral. Diabatic states are consistently denoted with lower case labels in order to differentiate them from adiabatic states such as S$_1$ and TT$_1$, which in turn consist of mixtures of diabatic states of different character. Beside the electronic degrees of freedom, the basis includes intramolecular vibrational states accompanying the electronic excitations. These are denoted $\tilde{\nu}$, $\bar\nu$, $\nu_+$, and $\nu_-$, for the singlet, triplet, cation and anion, respectively, and are always described in the vibrational potential pertaining to the electronic state of the molecule under consideration. In addition, the basis contains purely vibrational excitations, where $\nu$ vibrational quanta reside at a molecule in its singlet ground state (s$_0$).

The initial steps of singlet fission take place entirely within the singly-excited manifold. When considering linear absorption, or when characterizing the involved excited states, restricting the basis to this manifold alone suffices, as was done in I. However, when considering time-resolved spectroscopic experiments on fission materials, the ground state manifold and doubly-excited manifolds must be included as well. Similarly to I, we adapt the ``two-particle'' approximation,\cite{Philpott_71a, Spano_02a} such that the ground state manifold is spanned by the states
\begin{align}
\Ket{\BS{\T{s}_0}_m,\nu},
\end{align}
and
\begin{align}
\Ket{\BS{\T{s}_0}_m,\nu,\BS{\T{s}_0}_{m'},\nu'}.
\end{align}
Here the first state represents a purely vibrational excitation involving $\nu(\geq1)$ quanta in the s$_0$ potential of molecule $m$. In the second state, such a vibration is accompanied by a vibrational excitation involving $\nu'(\geq1)$ quanta located at site $m'$. (Note that $m>m'$ is imposed in order to avoid double counting.) Additionally, there is the unique state denoted $\Ket{\T{S}_0}$, in which all molecules are both electronically and vibrationally unexcited.

The doubly-excited manifold, on the other hand, consists of triplet pairs in which one member resides in a higher-lying state. Higher-lying states contributing to time-resolved spectroscopic measurements of singlet fission have been associated with the molecular excitations t$_2$ and t$_3$.\cite{Marciniak_07a, Marciniak_09a, Thorsmolle_09a, Thorsmolle_09b, Rao_10a, Rao_11a, Wilson_13a} We will nevertheless remain non-specific for the moment and consider for each set of calculations a single higher-lying state referred to as t$_n$, which suffices to account for the 2DES measurements, while leaving a discussion on the nature of this state to Sec.~\ref{Sec_TripletTransitions}. The doubly-excited triplet pairs are thus expressed as
\begin{align}
\Ket{\BS{\T{t}_1}_m,\bar\nu,\BS{\T{t}_n}_{m'},\bar{\bar\nu}'},
\end{align}
representing a t$_1$ excitation at molecule $m$, accompanied by $\bar\nu$ vibrational quanta, and a t$_n$ excitation at molecule $m'$ involving $\bar{\bar\nu}'$ quanta in the associated potential. Similarly to pair excitations in the singly-excited manifold, we assume this multi-exciton state to be entangled in a zero spin configuration, following the symmetry-adapted linear combination analogous to that defined for the (singly-excited) correlated triplet pair state in Ref.~\citenum{Berkelbach_13a}. The rationale behind this is that optical transitions are spin-conserving, and considering spin dephasing to be insignificant at the ultra-fast timescales considered in this work,\cite{Tasarov_97a} only such overall singlet states can be populated. In principle, the doubly-excited manifold also includes pairs of singly-excited basis states, which can become populated via two optical excitations. However, for crystalline pentacene, the associated spectroscopic signals do not benefit from the large dipole moment associated with the transition between triplets (as discussed in Sec.~\ref{Sec_TripletTDM}), and are therefore much weaker than the signals associated with the doubly-excited triplet pairs. Considering this, and the fact that such basis states would render the calculations prohibitively expensive, we have excluded such states from the doubly-excited manifold.

From the diabatic basis, adiabatic states are obtained upon solving the time-independent Schr\"odinger equation, $\Op{H}\Ket{\alpha}=\omega_\alpha\Ket{\alpha}$, where $\Op{H}$ denotes the Hamiltonian of the material under consideration, and $\Ket{\alpha}$ and $\omega_\alpha$ represent an (adiabatic) eigenstate and the associated eigenenergy, respectively. (Similarly to I, we apply $\hbar=1$ throughout this work.) Since we are concerned with three non-interacting excitation manifolds, the Hamiltonian can be subdivided according to
\begin{align}
\Op{H}=\Op{H}_0+\Op{H}_1+\Op{H}_2,
\end{align}
where the contributions with subscript 0, 1, and 2 act in the subspace corresponding to the ground state, singly-excited, and doubly-excited manifolds, respectively. For the ease of discussion, we will also explicitly differentiate between the eigenfunctions pertaining to these manifolds, denoting them as $\Ket{\alpha_0}$, $\Ket{\alpha_1}$, and $\Ket{\alpha_2}$, respectively. We refer to I for a formulation and discussion of the singly-excited Hamiltonian $\Op{H}_1$, which includes the couplings and energies associated with the vibronic s$_1$, t$_1$, c and a states, and continue to formulate $\Op{H}_0$ and $\Op{H}_2$.

Taking the energy of S$_0$ as a reference, the ground state Hamiltonian contains only the vibrational energy, $\Op{H}_0=\Op{H}_{\omega_0}$, given by
\begin{align}
\Op{H}_{\omega_0}=\omega_0\sum_m\Op{b}_m^\dagger\Op{b}_m.
\end{align}
Here, $\Op{b}_m$ and $\Op{b}_m^{\dagger}$ represent the vibrational ladder operators pertaining to the s$_0$ potential at molecule $m$, and $\omega_0$ is the vibrational quantum. The doubly-excited Hamiltonian, on the other hand, contains, in addition to vibrational energy, a contribution from electronic energy and a vibronic coupling term,
\begin{align}
\Op{H}_2=&\;E_{\T{t}_1\T{t}_n}\sum_{m\neq m'}\Ket{\BS{\T{t}_1}_m\BS{\T{t}_n}_{m'}}\Bra{\BS{\T{t}_1}_m\BS{\T{t}_n}_{m'}}+\Op{H}_{\omega_0}\\
&+\omega_0\!\!\!\sum_{x=\T{t}_1,\T{t}_n}\sum_m\;\lambda_x(\hat b_m^\dagger+\hat b_m+\lambda_x)\Ket{\BS{x}_m}\Bra{\BS{x}_m}.\nonumber
\end{align}
Here, $E_{\T{t}_1\T{t}_n}$ is the diagonal energy of the diabatic pair state involving the higher-lying triplet (assumed to be independent of the $m$ and $m'$ combination). In the vibronic coupling term, $x$ runs over the molecular excitations t$_1$ and t$_n$, and $\lambda_x$ is the associated vibronic coupling strength, the square of which corresponds to the Huang-Rhys (HR) factor. It should be noted that both $\Op{H}_0$ and $\Op{H}_2$ are diagonal operators in the diabatic basis, and consequently, the adiabatic states in the associated manifolds are identical to the diabatic ones. However, whereas intermolecular couplings in the ground state manifold are typically very weak, substantial couplings between doubly-excited states are plausible. The effect of such couplings is touched upon in Sec.~\ref{Sec_TripletTDM}.

\subsection{Two-dimensional electronic spectroscopy}\label{Sec_Spectroscopy}

Of central importance to spectroscopy, be it linear absorption or time-resolved optical techniques, is the transition dipole moment operator $\Op{\V{M}}$ of the material under consideration (bold notation is used throughout to refer to three-dimensional vector properties). Similarly to the Hamiltonian, we subdivide this operator into contributions associated with different (pairs of) excited state manifolds as
\begin{align}
\Op{\V{M}}=\Op{\V{M}}^{0-1}+\Op{\V{M}}^{1-2}.
\end{align}
The first term couples the ground state and the singly-excited manifolds through the optical transition from s$_0$ to s$_1$,
\begin{align}
\Op{\V{M}}^{0-1}=\sum_m\V{\mu}^{\T{s}_0-\T{s}_1}_m\Ket{\BS{\T{s}_0}_m}\Bra{\BS{\T{s}_1}_m}+\T{H.c.},
\end{align}
where $\V{\mu}^{\T{s}_0-\T{s}_1}_m$ is the associated transition dipole moment of molecule $m$. In contrast, the singly-excited manifold is coupled to the doubly-excited manifold through optical transitions between t$_1$ and t$_n$,
\begin{align}
\Op{\V{M}}^{1-2}=\sum_m\V{\mu}^{\T{t}_1-\T{t}_n}_m\Ket{\BS{\T{t}_1}_m}\Bra{\BS{\T{t}_n}_m}+\T{H.c.},
\end{align}
where $\V{\mu}^{\T{t}_1-\T{t}_n}_m$ is the corresponding transition dipole moment. Note that since $\Op{\V{M}}^{0-1}$ exclusively contributes to linear absorption, only this part of the transition dipole moment operator was considered in I. However, as depicted in Fig.~\ref{Fig_FissionScheme}, an evaluation of 2DES as presented here requires both contributions to $\Op{\V{M}}$ to be included.

Since we are concerned with one-component crystals (as opposed to cocrystals), all molecular transition dipole moments can be considered identical, aside from the description of the orientation. We can therefore set $\V{\mu}^{\T{s}_0-\T{s}_1}_m=\mu^{\T{s}_0-\T{s}_1}\V{e}^{\T{s}_0-\T{s}_1}_m$ and $\V{\mu}^{\T{t}_1-\T{t}_n}_m=\mu^{\T{t}_1-\T{t}_n}\V{e}^{\T{t}_1-\T{t}_n}_m$, where the unit vector $\V{e}^{\T{i}-\T{f}}_m$ represents the direction of the i$-$f transition dipole moment at molecule $m$, and $\mu^{\T{i}-\T{f}}$ is the associated ($m$-independent) magnitude. Furthermore, for the sake of clarity, it is helpful to additionally formulate transition dipole moment operators for which the magnitudes are divided out,
\begin{align}
\Op{\tilde{\V{M}}}^{0-1}=\sum_m\V{e}^{\T{s}_0-\T{s}_1}_m\Ket{\BS{\T{s}_0}_m}\Bra{\BS{\T{s}_1}_m}+\T{H.c.},
\end{align}
and
\begin{align}
\Op{\tilde{\V{M}}}^{1-2}=\sum_m\V{e}^{\T{t}_1-\T{t}_n}_m\Ket{\BS{\T{t}_1}_m}\Bra{\BS{\T{t}_n}_m}+\T{H.c.},
\end{align}
such that the original transition dipole moment operators are retained through $\Op{\V{M}}^{0-1}=\mu^{\T{s}_0-\T{s}_1}\Op{\tilde{\V{M}}}^{0-1}$ and $\Op{\V{M}}^{1-2}=\mu^{\T{t}_1-\T{t}_n}\Op{\tilde{\V{M}}}^{1-2}$.

\begin{figure}
\includegraphics{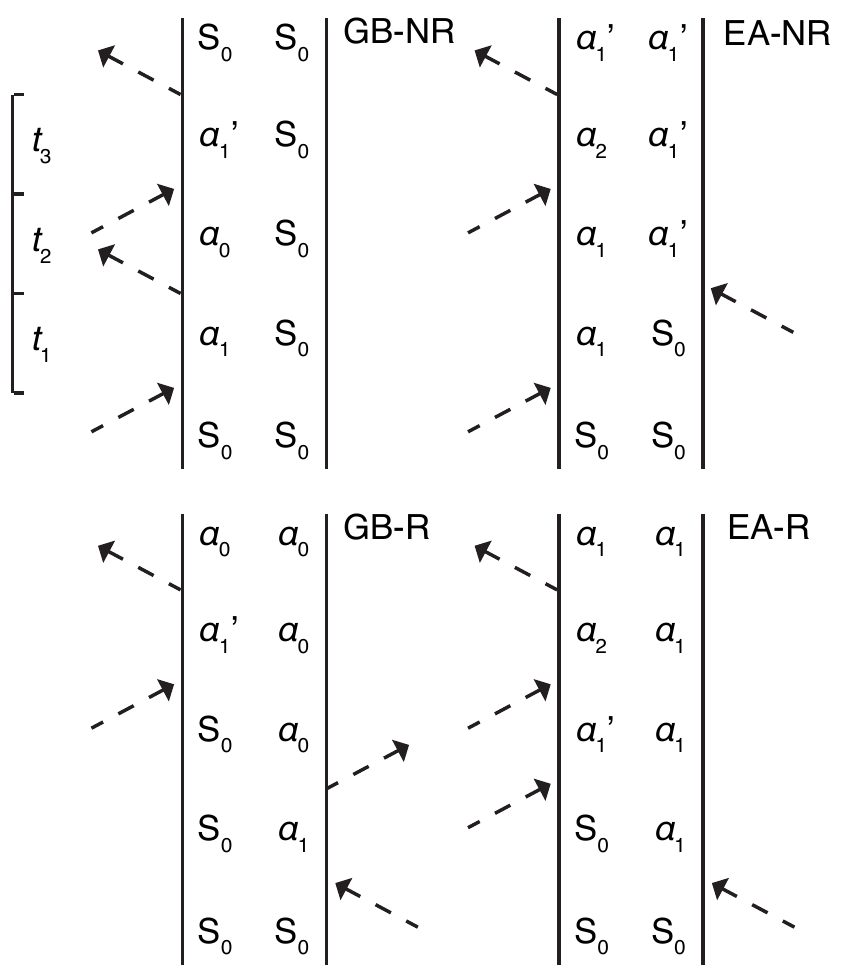}
\caption{Double-sided Feynman diagrams\cite{Mukamel_1, Jonas_03a, Cho_08a} describing the different sequences of the light-matter interactions underlying 2DES and transient absorption of fission materials. The diagrams are labeled according to their physical origin -- ground state bleach (GB) or excited state absorption (EA) -- and phase sequence -- non-rephasing (NR) or rephasing (R). For each diagram, the left (right) side describes the evolution of ket (bra) component of the molecular density matrix. Dashed arrows indicate light-matter interactions exciting (ingoing) and de-exciting (outgoing) the molecular system. For clarity, the ket and bra wavefunctions related to the ground, singly, and doubly excited manifolds are labeled with $\alpha_0$, $\alpha_1$, and $\alpha_2$, respectively (with or without prime). On the left, $t_1$, $t_2$, and $t_3$ indicate the ``pump'' time, waiting time, and ``probe'' time. }
\label{Fig_Diagrams}
\end{figure}

Whereas linear absorption is readily calculated via Fermi's Golden Rule, 2DES is more naturally evaluated in a time-resolved framework, through the calculation of the associated optical response functions.\cite{Mukamel_1, Jonas_03a, Cho_08a} These response functions derive from four light-matter interactions, separated in time by the intervals $t_1$, $t_2$, and $t_3$. The first two interactions constitute the aforementioned ``pump'' event, whereas the second interaction pair makes up the ``probe'' event, while the time interval $t_2$, the waiting time, is varied to dynamically study the molecular sample. What distinguishes 2DES \cite{Hybl_98a} (as well as its infrared analogue \cite{Hamm_98a}) from transient absorption is the ability to simultaneously resolve the pump and probe frequencies. Nevertheless, the underlying response functions are formally equivalent.

The response functions of interest to singlet fission can be divided into four essentially different forms, based on fundamentally different ways in which the density matrix is affected by the four optical interactions. These are conveniently represented by means of double-sided Feynman diagrams, describing the time-evolution of the molecular density matrix, as shown in Fig.~\ref{Fig_Diagrams}. The first difference lies in the relative phase accumulations during the pump and probe events, which could have equal signs (non-rephasing, NR) or opposite signs (rephasing, R). The second difference derives from the two physical processes that independently contribute to the spectral signal. One process, referred to as excited state absorption (EA), is light absorption during the probe event by molecules that have been promoted to the singly-excited manifold by the pump event. The other relates to the bleaching of ground state absorption during the probe event, due to molecules that have been excited by the pump event, which is referred to as ground state bleach (GB).

Under experimental conditions, it is fundamentally impossible to disentangle GB and EA signals, and the detected spectrum can at most be dissected into NR and R contributions. This is in marked contrast to numerical calculations, which proceed by separate evaluation of GB and EA, providing an easy venue for comparing the associated signals side-by-side. Singlet fission studies which used transient absorption \cite{Rao_10a, Burdett_10a, Wilson_11a, Rao_11a, Burdett_11a, Ramanan_12a, Roberts_12a, Musser_15a, Pensack_16a} or 2DES \cite{Bakulin_16a} have monitored triplet transitions that occur via EA as a dynamic probe of the fission process. From this perspective, GB is an undesirable contributor which complicates the interpretation of the measurements due to spectral congestion.

We further note that, in addition to EA and GB, a third process commonly contributes to time-resolved spectra, in which the probe event induces emission from singly-excited molecules. However, fission in bulk pentacene is known to depopulate the singlet state within $\sim$80~fs,\cite{Wilson_11a, Chan_11a} and radiative transitions from triplet pair states are generally very weak. As a result, such stimulated emission is not observed for this material,\cite{Rao_10a, Rao_11a} or at most at very early times,\cite{Wilson_11a} and is therefore neglected in our model. (For the same reasons, EA pathways involving transitions from the singlet state to higher-lying singlet excitations are neglected.)

The practical implementation of time-resolved spectroscopy relies on the realization of ultra-short laser pulses, particularly so for 2DES, which implies a broad laser spectrum in the frequency domain. Nevertheless, the spectral widths of fission materials typically exceed 1~eV, which is unreachable even by most 2DES setups. A simple method to account for a finite laser bandwidth is to multiply the transition dipole moment operator with an adiabatic window function, $\tilde{F}(\omega)$.\cite{Fuller_14a} Accordingly, the transition dipole moment operator is replaced by a laser-corrected analogue according to
\begin{align}
\Op{\tilde{\V{M}}}\leftarrow\sum_{\alpha,\alpha'}\Ket{\alpha}\Bra{\alpha}\Op{\tilde{\V{M}}}\tilde{F}\big(\vert\omega_\alpha-\omega_{\alpha'}\vert\big)\Ket{\alpha'}\Bra{\alpha'},
\label{Eq_LaserDipole}
\end{align}
where $\alpha$ and $\alpha'$ run over all manifolds. We have adopted this method, while following the example of Ref.~\citenum{Bakulin_16a} by using a discrete window function $\tilde{F}(\omega)=H(\omega-\omega_\T{L})H(\omega_\T{U}-\omega)$, where $H(\omega)$ is the Heaviside step function, and $\omega_\T{L}$ and $\omega_\T{U}$ characterize the lower and upper bounds of the laser spectrum, respectively. Although noting that more sophisticated approaches for incorporating the laser spectrum are available,\cite{Kjellberg_06a, Tempelaar_16a} we find the current approach sufficient for our purpose.

The response functions depicted in Fig.~\ref{Fig_Diagrams} are formulated as
\begin{widetext}
\begin{align}
R^\T{GB-NR}(t_1,t_2,t_3)=(\mu^{\T{s}_0-\T{s}_1})^4\!\!\sum_{j_1,j_2,j_3,j_4}\!\!A_{j_1,j_2,j_3,j_4}\Bra{\T{S}_0}\Op{\tilde{M}}^{0-1}_{j_4}\Op{U}(t_3)\Op{\tilde{M}}^{0-1}_{j_3}\Op{U}(t_2)\Op{\tilde{M}}^{0-1}_{j_2}\Op{U}(t_1)\Op{\tilde{M}}^{0-1}_{j_1}\Ket{\T{S}_0},
\label{Eq_Responses}
\end{align}
\begin{align}
R^\T{GB-R}(t_1,t_2,t_3)=(\mu^{\T{s}_0-\T{s}_1})^4\!\!
\sum_{j_1,j_2,j_3,j_4}\!\!A_{j_1,j_2,j_3,j_4}\Bra{\T{S}_0}\Op{\tilde{M}}^{0-1}_{j_1}\Op{U}^\dagger(t_1)\Op{\tilde{M}}^{0-1}_{j_2}\Op{U}^\dagger(t_2+t_3)\Op{\tilde{M}}^{0-1}_{j_4}\Op{U}(t_3)\Op{\tilde{M}}^{0-1}_{j_3}\Ket{\T{S}_0},\nonumber
\end{align}
\begin{align}
R^\T{EA-NR}(t_1,t_2,t_3)=-(\mu^{\T{s}_0-\T{s}_1})^2(\mu^{\T{t}_1-\T{t}_n})^2\!\!
\sum_{j_1,j_2,j_3,j_4}\!\!A_{j_1,j_2,j_3,j_4}\Bra{\T{S}_0}\Op{\tilde{M}}^{0-1}_{j_2}\Op{U}^\dagger(t_2+t_3)\Op{\tilde{M}}^{1-2}_{j_4}\Op{U}(t_3)\Op{\tilde{M}}^{1-2}_{j_3}\Op{U}(t_1+t_2)\Op{\tilde{M}}^{0-1}_{j_1}\Ket{\T{S}_0},\nonumber
\end{align}
\begin{align}
R^\T{EA-R}(t_1,t_2,t_3)=-(\mu^{\T{s}_0-\T{s}_1})^2(\mu^{\T{t}_1-\T{t}_n})^2\!\!
\sum_{j_1,j_2,j_3,j_4\!\!}A_{j_1,j_2,j_3,j_4}\Bra{\T{S}_0}\Op{\tilde{M}}^{0-1}_{j_1}\Op{U}^\dagger(t_1+t_2+t_3)\Op{\tilde{M}}^{1-2}_{j_4}\Op{U}(t_3)\Op{\tilde{M}}^{1-2}_{j_3}\Op{U}(t_2)\Op{\tilde{M}}^{0-1}_{j_2}\Ket{\T{S}_0}.\nonumber
\end{align}
\end{widetext}
Here, $j_1$, $j_2$, $j_3$, and $j_4$ label the four pulse polarizations. The minus sign appearing in the EA response functions, resulting from an uneven number of optical interactions at the bra of the molecular density matrix, is representative of the absorptive character of the associated spectral signals. All response functions are formulated in the frame of reference of the molecular crystal, and $\Op{\tilde{M}}^{0-1}_{j_1}$ refers to the $j_1$-polarized component of $\Op{\tilde{\V{M}}}^{0-1}$ (while the other transition dipole moment operators are labeled analogously). However, by having all pulse polarizations in the summation run independently over the unit vectors in the molecular frame ($x$, $y$, and $z$), and by weighing the contributions by the appropriate tensor values $A_{j_1,j_2,j_3,j_4}$,\cite{Hochstrasser_01a} we effectively consider an isotropic collective of molecular clusters. This is used as a proxy for a polycrystalline material, as discussed in Sec.~\ref{Sec_2DES}. Lastly, $\Op{U}(t)=e^{-i\Op{H}t}$ is the molecular propagator, accounting for the time-evolution of the sample in between the (impulsive) light-matter interactions. Note that $\Op{U}(t)\Ket{\T{S}_0}=I$ (identity matrix) is omitted. Fourier transforming the response functions over $t_1$ and $t_3$ yields the 2D spectra associated with waiting time $t_2$ as a function of the pump frequency $\omega_1$ and the probe frequency $\omega_3$.

\subsection{Parameters for pentacene}\label{Sec_Parms}

\begin{table}
\begin{center}
\begin{tabular}{l l l l}
\hline
\hline
Parameter & Symbol & Value & \\
\hline
s$_0-$s$_1$ energy & $E_{\T{s}_1}$ & 2.09~eV & (16\;890~cm$^{-1}$) \\
s$_0-$t$_1$t$_1$ energy & $E_{\T{t}_1\T{t}_1}$ & 1.75~eV & (14\;140~cm$^{-1}$) \\
s$_0-$t$_1$t$_n$ energy & $E_{\T{t}_1\T{t}_n}$ & 3.57~eV & (28\;825~cm$^{-1}$) \\
s$_0-$s$_1$ tdm & $\mu^{\T{s}_0-\T{s}_1}$ & 1.0 (arb.~u.) & \\
t$_1-$t$_n$ tdm & $\mu^{\T{t}_1-\T{t}_n}$ & $\sim$15.0 (arb.~u.) & \\
Vibr.~energy & $\omega_0$ & 0.17~eV & (1380~cm$^{-1}$) \\
Huang-Rhys factors & & & \\
\quad s$_0-$s$_1$ & $\lambda_{\T{s}_1}^2$ & 1.1 & \\
\quad s$_0-$a & $\lambda_{\T{a}}^2$ & 0.29 & \\
\quad s$_0-$c & $\lambda_{\T{c}}^2$ & 0.39 & \\
\quad s$_0-$t$_1$ & $\lambda_{\T{t}_1}^2$ & 1.1 & \\
\quad s$_0-$t$_n$ & $\lambda_{\T{t}_n}^2$ & 0 & \\
\hline
\hline
\end{tabular}
\caption{Parameters applied in our model (``tdm'' stands for transition dipole moment). For electronic couplings, see the supplementary material of I. Varying values of $\mu^{\T{t}_1-\T{t}_n}$, $\lambda_{\T{t}_1}^2$, and $\lambda_{\T{t}_n}^2$ are investigated in Sec.~\ref{Sec_TripletTDM}.}
\label{Tab_Parms}
\end{center}
\end{table}

Tab.~\ref{Tab_Parms} summarizes the most important parameters used in the present article, most of which have been discussed in I. We proceed to discuss only those parameters that specifically pertain to 2DES, and which have therefore not been presented in I. We note, however, that all of these parameters are concerned with t$_n$, the higher-lying triplet level included in our basis set. Such high-energy states are challenging to evaluate through computations based on first principles, and equally difficult to characterize spectroscopically. As a result, relatively little \textit{a priori} knowledge about them is available. We have therefore motivated the values adapted for these parameters based on the agreement of our calculations with 2DES measurements in Secs.~\ref{Sec_2DES} and \ref{Sec_FT}. Accordingly, the high-energy triplet pair state energy is taken to be $E_{\T{t}_1\T{t}_n}=3.57$~eV (28\;825~cm$^{-1}$). For the HR factors associated with t$_n$, an \textit{ad hoc} value of $\lambda_{\T{t}_n}^2=0$ is applied, while varying values of this quantity are explored in Sec.~\ref{Sec_TripletTDM}.

The direction and magnitude of the t$_1$ to t$_n$ transition dipole moments are determined based on the crystal structure of pentacene,\cite{Holmes_99a} similar to the procedure for the s$_0$ to s$_1$ transition in I. Whereas $\V{e}^{\T{s}_0-\T{s}_1}_m$ is oriented along the short axis of molecules $m$, $\V{e}^{\T{t}_1-\T{t}_n}_m$ is found to be oriented along the long axis instead\cite{Smith_10a, Anger_12a} and adapted accordingly. Regarding their magnitudes $\mu^{\T{s}_0-\T{s}_1}$ and $\mu^{\T{t}_1-\T{t}_n}$, we first note that these quantities solely act as constants contributing to the response functions given by Eq.~\ref{Eq_Responses}. Specifically, the constant contributing to the GB pathways is given by $(\mu^{\T{s}_0-\T{s}_1})^4$, whereas the equivalent for EA is $(\mu^{\T{s}_0-\T{s}_1})^2(\mu^{\T{t}_1-\T{t}_n})^2$. Hence, upon setting $\mu^{\T{s}_0-\T{s}_1}$ to an arbitrary value, the magnitude of the triplet transition effectively acts as a weighing of the EA signal relative to the GB analogue through its scaling to $(\mu^{\T{t}_1-\T{t}_n}/\mu^{\T{s}_0-\T{s}_1})^2$. For this reason, and in the light of a current inconclusiveness concerning the strength of the triplet transitions,\cite{Ashpole_74a, Hellner_72a, Pabst_08a, Smith_10a, Zimmerman_11a, Anger_12a, Smith_13a, Bakulin_16a} we use $\mu^{\T{t}_1-\T{t}_n}$ as a free parameter, expressed in terms of $\mu^{\T{s}_0-\T{s}_1}$, while leaving a thorough discussion of this quantity to Sec.~\ref{Sec_TripletTDM}.

In I, we evaluated the convergence of the photophysics of pentacene with respect to the applied crystal size, and found that the linear optical properties are well accounted for when limiting the crystal to 3$\times$3 unit cells extending in the crystallographic $ab$-plane (while imposing periodic boundary conditions). For pentacene, having 2 inequivalent molecules per unit cell, this amounts to 18 molecular sites. Still, the vibronic basis set employed in our model scales rather unfavorably with the number of electronic degrees of freedom, yielding a total of 2430 states, despite an optimal truncation of the basis set. Simulations of 2DES involving this number of states are just within the computational capacities at our disposal, and we have adapted the crystal size accordingly. In doing so, the diabatic bases spanning the ground state and doubly-excited state manifolds are truncated analogously to the singly-excited basis discussed in I. Specifically, the maximum t$_1$ to t$_n$ separation is set to 6.2 \AA, the separation between the two-particle states in the ground state manifold is limited to 20 \AA, and the total number of vibrational quanta per basis state is limited to 2.

\section{Results and discussion}\label{Sec_Results}

\subsection{Static 2D spectra}\label{Sec_2DES}

Linear absorption has proven indispensable to validate microscopic models of singlet fission.\cite{Yamagata_11a, Beljonne_13a, Berkelbach_14a, Hestand_15a} Nevertheless, with the recently reported 2DES measurements on polycrystalline pentacene, the opportunity to subject models to an even more stringent test arises. The ability of 2DES to project the optical response of a molecular material to an additional spectral dimension allows one to directly resolve the correlations between excitations over time. On the one hand, non-oscillatory signals are representative of energy transfer pathways and couplings between excited states.\cite{Mukamel_1, Jonas_03a, Brixner_05a, Cho_08a} On the other hand, oscillatory signals represent coherent dynamics.\cite{Mukamel_1, Jonas_03a, Engel_07a, Cho_08a} In 2DES experiments on pentacene, the optical transition between the ground state and the correlated triplet pair state has been directly detected with the help of such quantum beats.\cite{Bakulin_16a} Before turning our attention to this spectroscopic result, we first focus on the non-oscillatory signals by comparing our simulations to the measurements of static 2D spectra.

In calculating the static 2D spectra, the underlying response functions (Eq.~\ref{Eq_Responses}) are evaluated while scanning the pump ($t_1$) and probe ($t_3$) times from 0 to 200~fs using a 1~fs time resolution. As a convenient way to realize spectral broadening, a multiplication by the exponential $e^{-(t_1+t_3)/\tau}$ is carried out before performing the 2D Fourier transforms, resulting in Lorentzian peak profiles along $\omega_1$ and $\omega_3$. In doing so, the lifetime parameter $\tau$ is adjusted such that the Lorentzian width corresponds to 27 meV (214~cm$^{-1}$). While evaluating the crystal, a weighted average is taken over all possible polarization sequences, as embodied by Eq.~\ref{Eq_Responses}, effectively regarding the crystal as a member of an isotropic ensemble. In I, this approach was justified for linear absorption of polycrystalline pentacene, since the molecular s$_0-$s$_1$ transitions are polarized in the $ab$-plane, for which an effective isotropy occurs in the polycrystalline state. However, the t$_1-$t$_n$ transition is oriented parallel to the crystallographic $c$-axis, a direction along which polycrystalline pentacene has been shown to be reasonably ordered.\cite{Ruiz_04a, Nickel_04a} Nevertheless, the 2DES measurements were recorded by orienting the laser beams at a 50$^\circ$ angle relative to this axis,\cite{Bakulin_16a} in order to compromise between maximizing the t$_1-$t$_n$ signal and minimizing the strong absorption cross-section in the $ab$-plane (due to extended in-plane crystal sizes).\cite{Marciniak_07a, Marciniak_09a, Wilson_13a} We argue that the isotropic calculations presented here, by its weighing of both weak in-plane and strong out-of-plane t$_1-$t$_n$ signals, forms a reasonable proxy for such an experimental setup. Furthermore, we point out that the 2DES spectral features are, by construction, qualitatively insensitive to such polarization details, with quantitative differences anticipated only in the relative intensities of GB and EA signals.

Another factor that has only a quantitative impact on the features observed in 2DES is the magnitude of the triplet transition dipole moment, $\mu^{\T{t}_1-\T{t}_n}$, relative to that of the singlet transition. As argued in Sec.~\ref{Sec_Parms}, this quantity effectively modulates the EA:GB signal ratio via its $(\mu^{\T{t}_1-\T{t}_n})/(\mu^{\T{s}_0-\T{s}_1})^2$ scaling. We accordingly utilize the magnitude of this quantity as a parameter to reach agreement with experiment regarding the weighing of oscillatory spectral features from GB and EA in Sec.~\ref{Sec_FT}, yielding $\mu^{\T{t}_1-\T{t}_n}\sim15\mu^{\T{s}_0-\T{s}_1}$. This value is significantly larger than the magnitude of $2.5\mu^{\T{s}_0-\T{s}_1}$ used in the 2DES simulations in Ref.~\citenum{Bakulin_16a}, a difference that originates primarily from the much weaker TT$_1$ oscillator strength obtained in I, which was found to be more realistic with respect to linear absorption features. In Sec.~\ref{Sec_TripletTDM}, we highlight several physical factors that additionally impact the EA signal strength, and which allow to account for the 2DES measurements using smaller values of $\mu_{\T{t}_1-\T{t}_n}$. Finally, we note that the quantitative accuracy of our modeling of the GB and EA signals is inherently limited, since aspects such as dynamical population of TT$_1$ and realistic pulse shapes are not accounted for.

\begin{figure}
\includegraphics{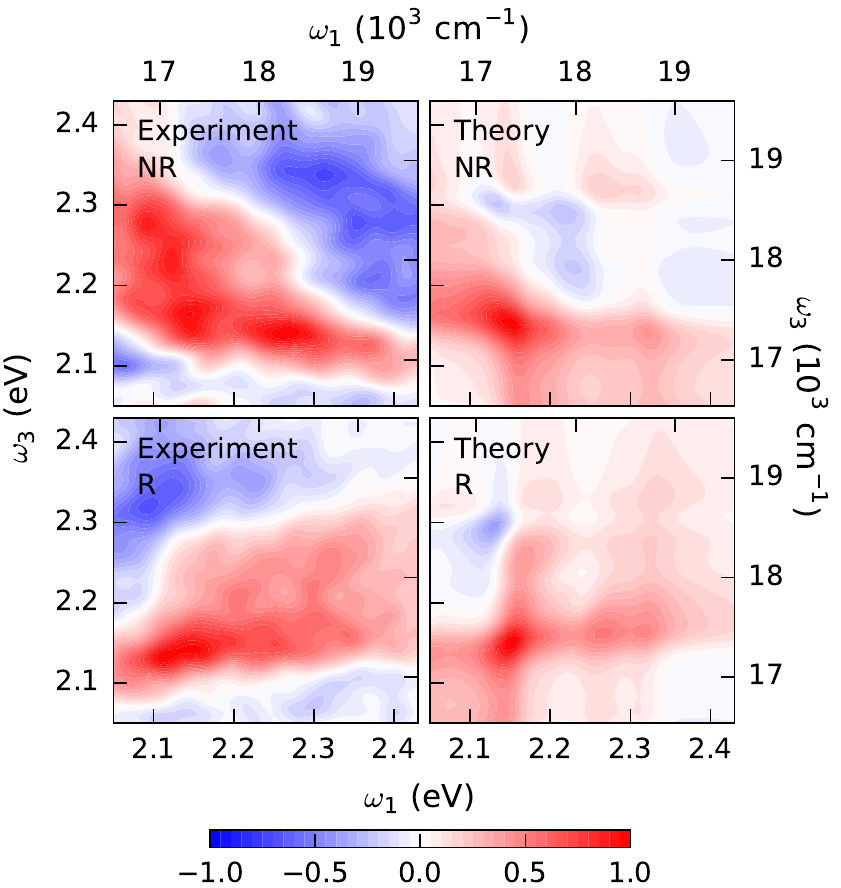}
\caption{Static 2DES in the high-energy spectral region of polycrystalline pentacene. Shown are measured (left panels) and calculated (right panels) real-valued 2D spectra at an 80~fs waiting time. Spectra obtained through the non-rephasing (NR) and rephasing (R) detection scheme are shown in the top and bottom row, respectively. The spectra are normalized individually.}
\label{Fig_HighEnergy}
\end{figure}

Before focusing on the spectral region near the energy of TT$_1$, we first compare static 2DES calculated through our model with measurements in the higher-energy region where multiple bright exciton states render congested spectra, making for a challenging test case. Shown in Fig.~\ref{Fig_HighEnergy} are experimental real-valued 2D spectra from Ref.~\citenum{Bakulin_16a}. The spectra were recorded in the NR and R phase sequences, at a waiting time $t_2=80$~fs, and using a laser bandwidth ranging from $\omega_{\T{L}}=2.05$~eV (16\;500~cm$^{-1}$) to $\omega_{\T{U}}=2.36$~eV (19\;000~cm$^{-1}$). By setting the window function (Eq.~\ref{Eq_LaserDipole}) accordingly, we obtain the calculated results also shown in Fig.~\ref{Fig_HighEnergy}. Since the t$_1-$t$_n$ transition lies outside this window, we instead included for these higher-energy region calculations a higher-energy t$_{n'}$ ($n'>n$) state using $E_{\T{t}_{n'}}=4.44$~eV (35\;800~cm$^{-1}$), based on a comparison with the experimental 2D spectra. Despite not being otherwise parametrized to reproduce the 2D spectra, our model reaches a very high level of agreement, reproducing the dominating diagonal peak at 2.15~eV present in both the NR and R spectra, as well as the associated cross-peak at probe energy $\omega_3=2.25$~eV and the series of cross-peaks extending over the range of pump energies from $\omega_1=2.25$~eV up to 2.35~eV. The main discrepancies concern the spectral fine structure and the intensity of negative signal. Based on our comparison between linear absorption bands calculated for a 3$\times$3 and a 6$\times$6 pentacene crystal in I, the former can be attributed to the finite crystal dimension used in the 2DES calculations, for which the fine structure in the high-energy spectral region has not fully converged. The lack of negative signal, on the other hand, is mostly attributable to the absence of fission dynamics in our model, which, when included, would considerably enhance static EA contributions through a buildup of triplet product states. Furthermore, the negative features appearing in the experimental spectra at a detection energy close to $\omega_{\T{L}}$ are likely artifacts due to the laser pulse that go beyond the windowing effect applied through Eq.~\ref{Eq_LaserDipole}, being consistent with theoretical predictions from Ref.~\citenum{Tempelaar_16a} (see Fig.~2 in that reference). We note, however, that most of the features are reproduced at least qualitatively.

\begin{figure}
\includegraphics{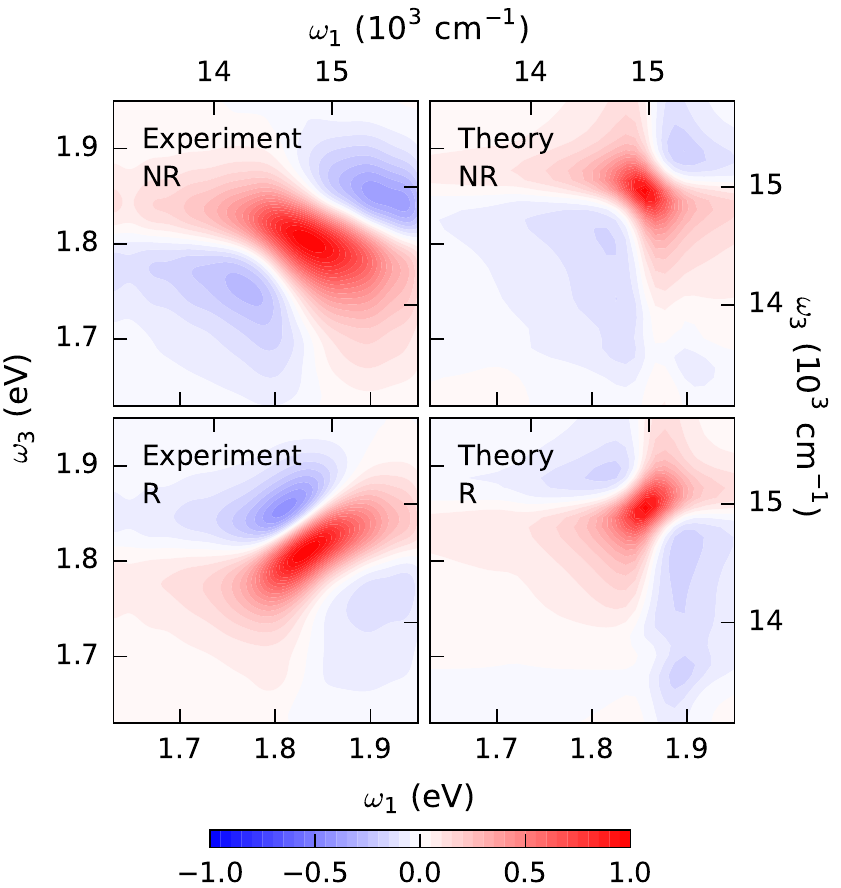}
\caption{Static 2DES of polycrystalline pentacene in the region covering the optical transitions from S$_0$ to S$_1$ and TT$_1$. Shown are 2D spectra at a waiting time of 80~fs. The presentation is analogous to Fig.~\ref{Fig_HighEnergy}.}
\label{Fig_LowEnergy}
\end{figure}

A similar comparison is shown in Fig.~\ref{Fig_LowEnergy} for the 2D spectra recorded at 80~fs in the region from $\omega_{\T{L}}=1.61$~eV (12\;950~cm$^{-1}$) to $\omega_{\T{U}}=1.96$~eV (15\;800~cm$^{-1}$), which covers the optical transition between the S$_0$ and TT$_1$. In marked contrast to the higher-energy equivalents, both the static NR and R spectra show very little structure apart from a pronounced diagonal peak located at 1.85~eV and associated with S$_1$. When compared to our calculated results, this peak is somewhat red-shifted in the measurements, which is a known distortion owing to the finite pulse bandwidth.\cite{Tempelaar_16a} Regardless, both the experimental and numerical spectra show no discernible features around $\omega_1=1.72$~eV, which corresponds to the energy of TT$_1$, implying that the S$_0-$TT$_1$ transition between these states is indiscernible through static 2D spectra. However, as was first demonstrated in Ref.~\citenum{Bakulin_16a}, this transition can nonetheless be unveiled by Fourier transforming the oscillatory 2D signal relative to the waiting time, yielding Fourier transform amplitude maps.

\subsection{Fourier transform amplitude maps}\label{Sec_FT}

Numerical Fourier transform (FT) amplitude maps are derived from 2D spectra calculated at 5~fs intervals of the waiting time $t_2$ between 0 and 1500~fs. In order to keep the computational costs manageable, we have truncated $t_1$ and $t_3$ to 100~fs, which restricts the resolution of the conjugate energies $\omega_1$ and $\omega_3$, but which is still sufficient to produce satisfactory spectral results. For each ($\omega_1,\omega_3$) pair, complex-valued time traces are composed by taking the associated $t_2$-dependent signal from the 2D spectra. These traces are fitted to an exponential, whereupon a fast Fourier transform is applied to the residue, the result of which forms a complex-valued frequency spectrum. The absolute value of this spectrum at a certain frequency $\omega_2$, added to its analogue at $-\omega_2$, yields the FT amplitude map associated with $\omega_2$ as a function of $\omega_1$ and $\omega_3$.

\begin{figure}
\includegraphics{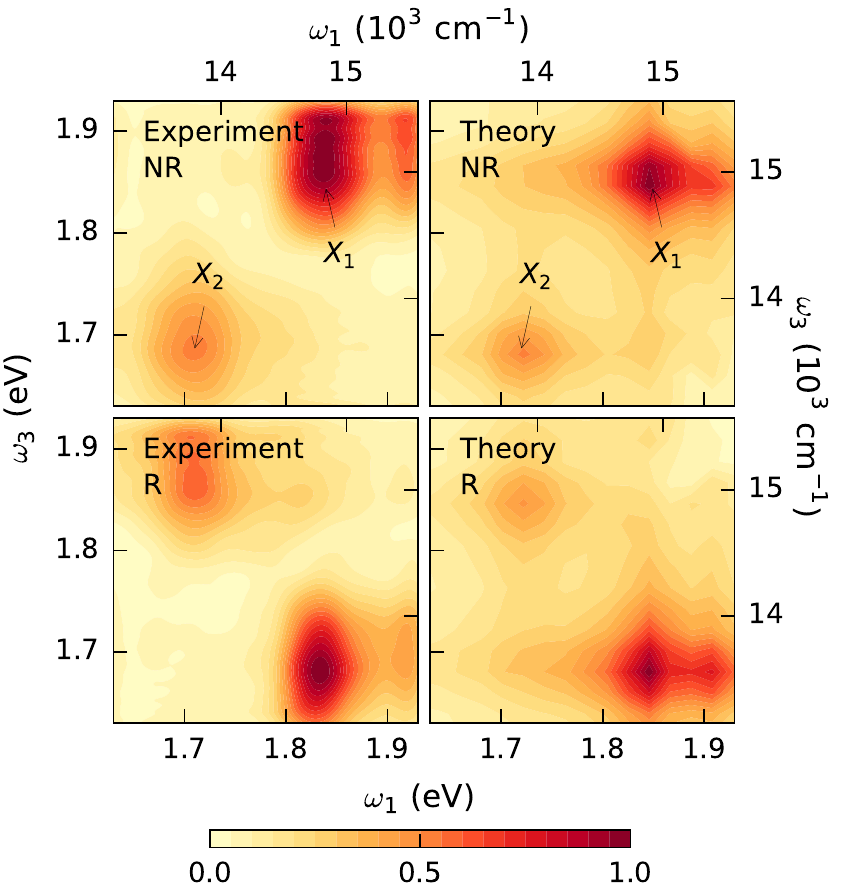}
\caption{Fourier transform amplitude maps taken at $\omega_2=169$~meV, derived from the non-rephasing (top row) and rephasing (bottom-row) 2D spectra in the region covering the optical transitions from S$_0$ to S$_1$ and TT$_1$. Measured results are shown on the left, whereas the calculated equivalents are shown on the right. All maps are normalized individually.}
\label{Fig_FTMaps}
\end{figure}

The experimental Fourier transform amplitude maps from Ref.~\citenum{Bakulin_16a} in the spectral region of TT$_1$ are shown in Fig.~\ref{Fig_FTMaps}. The associated Fourier transform frequency corresponds to $\omega_2=169$~meV (1360~cm$^{-1}$), which is approximately the symmetric stretching vibration $\omega_0$. Interestingly, the signals resolved upon Fourier transforming the optical response with respect to $t_2$ are obviously much richer in features than the static spectra shown in Fig.~\ref{Fig_LowEnergy} would suggest. In the NR map, the features are roughly concentrated in two peaks in the diagonal region, labeled $X_1$ and $X_2$ in order of descending pump energies. Whereas $X_2$ appears as a single peak, $X_1$ seems to be composed of a strong peak at the diagonal overlapping a weaker feature slightly blue-shifted along $\omega_1$. In the R map, on the other hand, two peaks are observable at off-diagonal locations. In both phasing sequences, the peak intensity at higher $\omega_1$ amounts to approximately three times the intensity of the lower $\omega_1$ peak.

\begin{figure}
\includegraphics{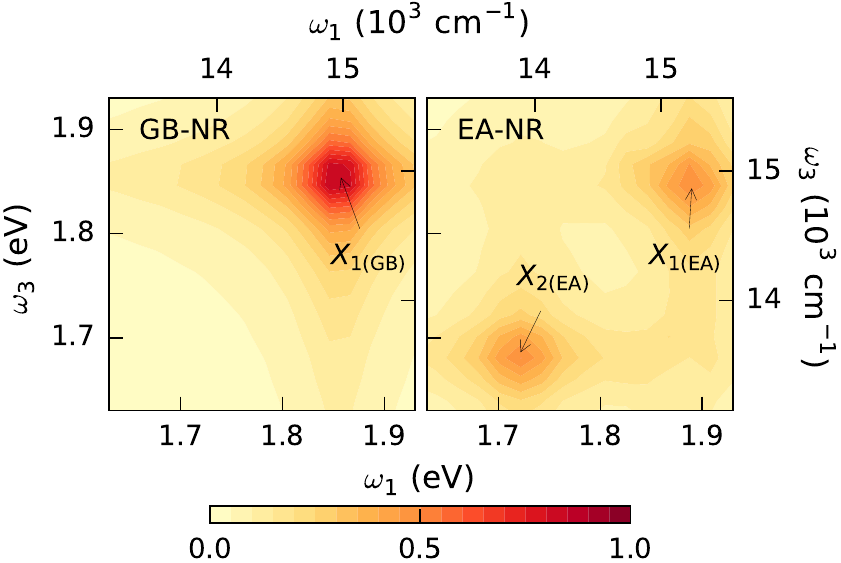}
\caption{Calculated Fourier transform amplitude maps at $\omega_2=169$~meV, derived from non-rephasing ground state bleach (left) and excited state absorption (right) signals. Both are normalized to the calculated non-rephasing map shown in Fig.~\ref{Fig_FTMaps}.}
\label{Fig_FTMapsDissect}
\end{figure}

Before turning our attention to the total NR and R maps resulting from our model, we first consider in Fig.~\ref{Fig_FTMapsDissect} the separate contributions from GB and EA to the NR oscillatory signal, which are inseparable under experimental conditions. These data substantiate that $X_1$ has two contributors, originating from GB and EA, whereas $X_2$ consists of a single feature owing to EA. Moreover, since the EA signal is modulated by the triplet transition dipole moment through its scaling to $(\mu^{\T{t}_1-\T{t}_n}/\mu^{\T{s}_0-\T{s}_1})^2$, agreement in the relative intensities of $X_1$ and $X_2$ can be reached with the measurements by adjusting this parameter in our model. The matching of these intensities motivates $\mu^{\T{t}_1-\T{t}_n}\sim15\mu^{\T{s}_0-\T{s}_1}$. Using this value, good overall agreement is reached for both the total R and NR maps, as can be seen in Fig.~\ref{Fig_FTMaps}. Nevertheless, we highlight in Sec.~\ref{Sec_TripletTDM} various molecular properties that additionally impact the $X_1$ and $X_2$ peak intensities, while leaving the oscillatory features qualitatively unaffected. By means of these properties, agreement for the FT amplitude maps can be reached using  different values of $\mu^{\T{t}_1-\T{t}_n}$.

\begin{figure}
\includegraphics{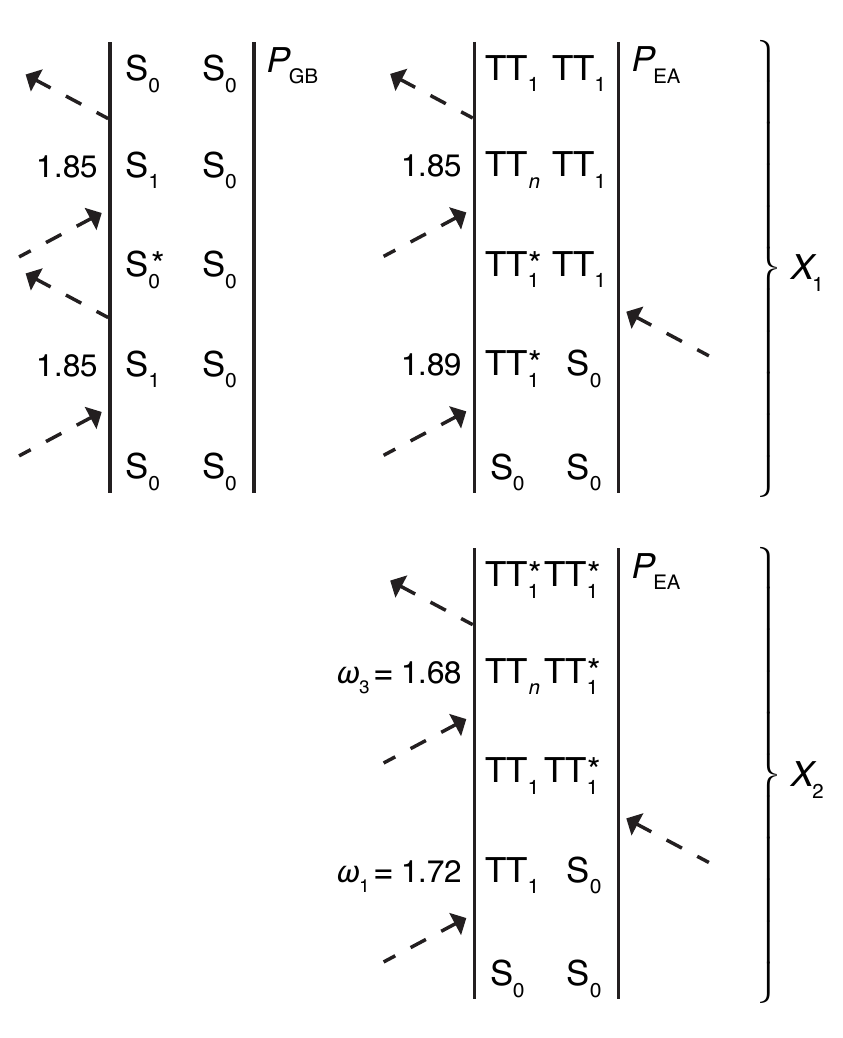}
\caption{Non-rephasing double-sided Feynman diagrams underlying peaks $X_1$ and $X_2$ in the Fourier transform amplitude maps associated with $\omega_2=169$~meV. The involved states are labeled in accordance with Fig.~\ref{Fig_FissionScheme}, including the vibrationless ground state S$_0$, whereas S$_0^*$ refers to electronic ground state dressed with a single vibration. Numbers indicate the energy differences (in eV) between the ket and bra states, which dictate the $\omega_1$ (bottom) and $\omega_3$ (top) energies at which the associated peaks appear in the maps.}
\label{Fig_DiagramsSpec}
\end{figure}

The spectral features observed in the GB and EA maps shown in Fig.~\ref{Fig_FTMapsDissect} are readily characterized based on the associated diagrams from Fig.~\ref{Fig_Diagrams} combined with the excited state energies of crystalline pentacene as determined in I. These Feynman pathways are shown in Fig.~\ref{Fig_DiagramsSpec}. We note that in addition to the vibrationless ground state S$_0$, the relevant diagrams include a single-quantum purely vibrational excitation. Denoting this excitation as S$_0^*$, we point out that it represents a manifold of (degenerate) states which were referred to as $\Ket{\BS{\T{s}_0}_m,\nu=1}$ in Sec.~\ref{Sec_Theory}. The features observed in the R maps are similarly accounted for based on the diagrams, upon inverting for each diagram the first two light-matter interactions, and these are therefore not explicitly considered here. The characterization shown in Fig.~\ref{Fig_DiagramsSpec} bears strong similarity with the analysis presented in Ref.~\citenum{Bakulin_16a} based on phenomenological modeling, although we note that the EA contribution to peak $X_1$ was not identified in this earlier work. As in Ref.~\citenum{Bakulin_16a}, we identify $X_2$ as a direct probe of the optical transition between S$_0$ and TT$_1$ as it maps the associated energy difference onto the pump axis, and is free of overlapping GB contributions.

Concordant with Ref.~\citenum{Bakulin_16a}, we find that the transition between S$_0$ and TT$_1$ exposes weak oscillations that are observed in the FT amplitude maps, in marked contrast to the non-oscillatory (static) signals which show no trace of this transition. In addition, note that in I we found the contribution of the S$_0-$TT$_1$ transition to linear absorption to be negligible. In order to better understand these differences, we proceed to examine the strength of the oscillatory signal at $X_2$, through which the S$_0-$TT$_1$ transition is detected. In accordance with Fig.~\ref{Fig_DiagramsSpec}, this strength is proportional to
\begin{align}
X_2\propto\;&\Bra{\T{S}_0}\Op{M}_{j_1}\Ket{\T{TT}_1}\Bra{\T{TT}_1}\Op{M}_{j_2}\Ket{\T{TT}_n}\nonumber\\
&\times\Bra{\T{TT}_n}\Op{M}_{j_3}\Ket{\T{TT}_1^*}\Bra{\T{TT}_1^*}\Op{M}_{j_4}\Ket{\T{S}_0},
\label{Eq_P_EA_Approx}
\end{align}
disregarding the polarization details for the moment. The first term in this product involves the transition dipole moment between S$_0$ and TT$_1$, which was found to be small in I. However, the remainder of the involved transition intensities may be sizable. The last term couples S$_0$ and the vibrationally dressed correlated triplet pair state TT$_1^*$, which, in I, was found to have a transition dipole moment whose magnitude is approximately one third of that of the bright S$_0-$S$_1$ transition. The remaining terms involve bright transitions between triplets. This transition dipole moment sequence yields much stronger signals than the imprint of TT$_1$ on linear absorption, given by $\vert\Bra{\T{S}_0}\Op{M}_{j_1}\Ket{\T{TT}_1}\vert^2$, which involves two weighings with the weak S$_0-$TT$_1$ transition dipole moment, and which does not benefit from the involvement of additional bright transitions. Finally, since the triplet transitions play an essential role in the prominence of $X_2$, and the spectroscopic detection of the correlated triplet pair state in general, we dedicate the following two sections to a more complete discussion of their nature.

\subsection{Identifying optical transitions between triplet states}\label{Sec_TripletTransitions}

Among the open questions related to singlet fission, the attribution of optical transitions between triplet excitations is perhaps the most controversial.\cite{Hellner_72a, Ashpole_74a, Marciniak_07a, Pabst_08a, Marciniak_09a, Thorsmolle_09a, Thorsmolle_09b, Rao_10a, Smith_10a, Rao_11a, Wilson_13a, Smith_13a, Bakulin_16a} As mentioned in Sec.~\ref{Sec_Theory}, studies have suggested two different higher-lying triplet excitations to be involved in time-resolved spectroscopic observations of singlet fission. In transient absorption measurements on polycrystalline pentacene, an EA feature centered at a probe energy of 2.34~eV has been attributed to the transition from t$_1$ to t$_3$,\cite{Marciniak_07a, Marciniak_09a, Rao_11a, Wilson_13a} whereas a broad feature ranging from 1.24 to 1.77~eV has been associated with the t$_1$ to t$_2$ transition.\cite{Thorsmolle_09a, Thorsmolle_09b, Rao_10a, Wilson_13a} This attribution is more or less supported by the energetics obtained through approximate coupled-cluster singles and doubles (CCSD), which predicted the t$_1-$t$_2$ and t$_1-$t$_3$ transitions to be located at 1.41~eV and 2.67~eV, respectively.\cite{Pabst_08a} However, whereas the latter was calculated to have a rather strong oscillator strength, the former turned out to be essentially dark, which is consistent with measurements on pentacene in the gas phase\cite{Ashpole_74a} and in solution.\cite{Hellner_72a} Whether crystallization of pentacene can alter the triplet transitions to the extent that such could explain these apparent differences has remained unclear.\cite{Smith_10a, Smith_13a} Interestingly, a recent study\cite{Khan_17a} suggested that local distortions of the crystal structure in pentacene is a possible source of the apparent enhancement in t$_1-$t$_2$ absorption observed in experiments. Using simulations of EA signals based on multiple reference singles and doubles configuration interactions (MRSDCI), the resulting t$_1-$t$_2$ oscillator was found to amount to roughly a quarter of the t$_1-$t$_3$ oscillator strength, while the associated transition energies were determined to be 1.80~eV and 2.36~eV, respectively.\cite{Khan_17a}

The attribution of triplet transitions based on experimental transient absorption spectra is complicated by overlapping GB features. This complication is largely overcome by 2DES via its potential to simultaneously resolve pump and probe energies. In the 2DES measurements on polycrystalline pentacene, two static EA features were identified at pump energies of 1.70 and 1.88~eV, growing in intensity to become resolvable at a waiting time of 150 fs.\cite{Bakulin_16a} These were associated with the triplet transitions involving t$_2$ and t$_3$, respectively, despite energetic differences relative to the aforementioned calculations and transient absorption measurements.

Although further experimental research in combination with first principles modeling is needed to resolve this controversy beyond a shadow of a doubt, our theoretical results provide useful information to facilitate this, especially in its reproduction of the experimental FT amplitude maps. In order to reach agreement for these maps, the doubly-excited manifold in our model is accounted for by considering a single kind of t$_n$ state, taking part in the pairs t$_1$t$_n$ that are parametrized by the associated energy $E_{\T{t}_1\T{t}_n}=3.57$~eV. As a consequence, TT$_n$ lies 1.85~eV above TT$_1$, which is close to the t$_1-$t$_2$ transition energy calculated through MRSDCI.\cite{Khan_17a} Consequentially, t$_n$ is naturally identified with t$_2$ (and TT$_n$ with TT$_2$). Moreover, TT$_n$ happens to lie 1.68~eV above the TT$_1^*$ energy as determined in I, hence, triplet transitions occur at 1.85~eV as well as 1.68~eV. We thus find a possible explanation for both static EA features observed in the 2DES measurements (within spectroscopic accuracy) based on t$_2$ alone.

We furthermore point out that the triplet transition at 1.85~eV overlaps with the absorption signal of S$_1$, which renders it difficult to resolve using transient absorption. Instead, the transition from TT$_1^*$ to TT$_n$ is expected to dominate the signal probed using this technique, which explains why the t$_1-$t$_2$ transition is observed at significantly lower energies. We note in closing that the transitions associated with t$_3$ are predicted to lie outside the spectral window of the FT amplitude maps, and therefore our analysis does not provide conclusive insights into these transitions. However, based on energetics, it is plausible that this state corresponds to t$_{n'}$ used to reproduce the 2D spectra at higher energies in Sec.~\ref{Sec_2DES}.

\subsection{Triplet transition dipole moment}\label{Sec_TripletTDM}

Regardless the nature of t$_n$, its substantial transition dipole moment relative to t$_1$ is essential to the optical detection of the S$_0-$TT$_1$ transition, as elaborated in Sec.~\ref{Sec_FT}. More specifically, the intensity of peak $X_2$ in the FT amplitude maps, through which this transition is detected, scales as $(\mu^{\T{t}_1-\T{t}_n})^2$ owing to its EA origin. Peak $X_1$, on the other hand, results from EA and GB and has therefore a different (and weaker) dependence on the triplet transition dipole moment. A comparison of both peak intensities in Sec.~\ref{Sec_FT} yielded $\mu^{\T{t}_1-\T{t}_n}\sim15\mu^{\T{s}_0-\T{s}_1}$, when applied in the framework of our model. This value compares reasonably well to CCSD calculations of $\mu^{\T{t}_1-\T{t}_3}$ (10.24 Debye, Ref.~\citenum{Pabst_08a}) and TD-DFT calculations of $\mu^{\T{s}_0-\T{s}_1}$ (1.16 Debye, Ref.~\citenum{Zimmerman_11a}), while being somewhat large compared to estimates of $\mu^{\T{t}_1-\T{t}_2}$, which even in the presence of crystal disorder is expected to amount to roughly a half of $\mu^{\T{t}_1-\T{t}_3}$.\cite{Khan_17a} Furthermore, a significantly smaller value, $\mu^{\T{t}_1-\T{t}_n}=2.5\mu^{\T{s}_0-\T{s}_1}$, was used in the phenomenological modeling of 2DES presented in Ref.~\citenum{Bakulin_16a}, for reasons discussion in Sec.~\ref{Sec_2DES}. This is illustrative of the uncertainty regarding the relative values of $\mu^{\T{s}_0-\T{s}_1}$ and $\mu^{\T{t}_1-\T{t}_n}$ in the literature.\cite{Ashpole_74a, Hellner_72a, Pabst_08a, Smith_10a, Zimmerman_11a, Anger_12a, Smith_13a, Bakulin_16a} For this reason, we bring forward several material properties which, alongside $\mu^{\T{t}_1-\T{t}_n}$, impact the $X_1$ and $X_2$ peak intensities, and through which the experimental FT amplitude maps can be numerically accounted for using different values of $\mu^{\T{t}_1-\T{t}_n}$. This exploration is by no means intended as an accurate determination of $\mu^{\T{t}_1-\T{t}_n}$, for which the unknowns in the material properties under consideration and the approximations made in our modeling are too restrictive. Rather, it serves to guide ongoing research by highlighting various factors that have to be taken into account when interpreting time-resolved spectroscopic measurements of singlet fission.

\begin{figure}
\includegraphics{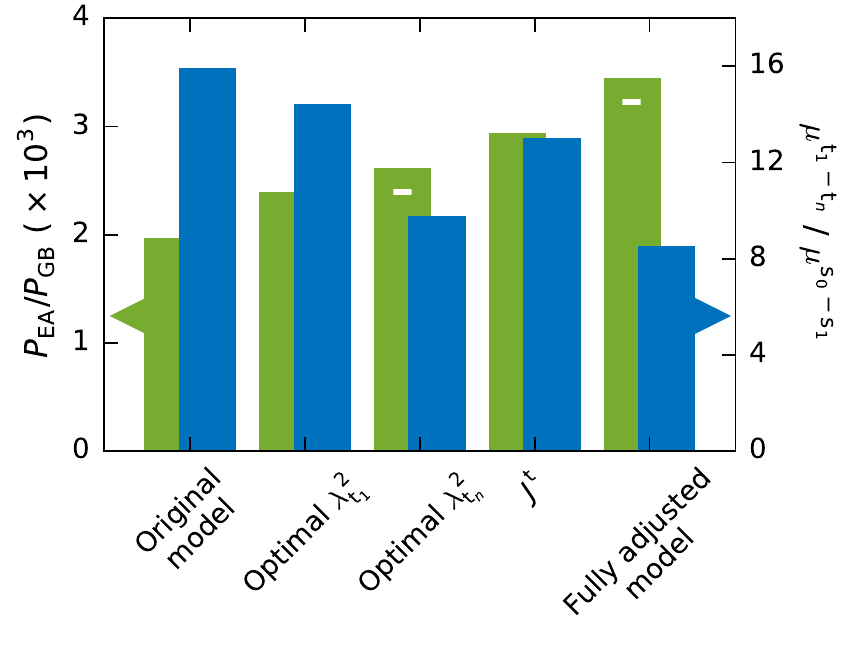}
\caption{Ratio of the coefficients quantifying the EA and GB oscillatory contributions ($P_\T{EA}/P_\T{GB}$, green bars), for which the transition dipole moments are disregarded, and the corresponding value of $\mu^{\T{t}_1-\T{t}_n}$ (in terms of $\mu^{\T{s}_0-\T{s}_1}$) required to reproduce the Fourier transform amplitude maps shown in Fig.~\ref{Fig_FTMaps} through Eq.~\ref{Eq_Peaks} (blue bars). Results calculated in the original model (outlined in Sec.~\ref{Sec_Theory}) are compared with modified models in which the t$_1$ and t$_n$ Huang-Rhys factors are optimized to maximize the EA:GB peak intensity, in which dipole-dipole couplings between triplet transitions are included, and where all of the above have been applied. Negative values of $P_\T{EA}/P_\T{GB}$ are denoted with a minus sign.}
\label{Fig_FTMapAnalysis}
\end{figure}

Instead of performing expensive simulations of FT amplitude maps based on 2D spectra, we proceed to use an approximate measure for the $X_1$ and $X_2$ peak intensities. First, the GB and EA signal strengths without inclusion of the transition dipole moment magnitudes are assessed by means of the coefficients
\begin{align}
P_\T{GB}=&\sum_{j_1,j_2,j_3,j_4}A_{j_1,j_2,j_3,j_4}\sum_{\alpha_0\in\T{S}_0^*}\Bra{\T{S}_0}\Op{\tilde{M}}_{j_1}\Ket{\T{S}_1}\nonumber\\
&\times\Bra{\T{S}_1}\Op{\tilde{M}}_{j_2}\Ket{\alpha_0}\Bra{\alpha_0}\Op{\tilde{M}}_{j_3}\Ket{\T{S}_1}\Bra{\T{S}_1}\Op{\tilde{M}}_{j_4}\Ket{\T{S}_0}
\label{Eq_P_GB}
\end{align}
and
\begin{align}
P_\T{EA}=&-\!\!\!\sum_{j_1,j_2,j_3,j_4}A_{j_1,j_2,j_3,j_4}\!\!\!\!\!\!\!\!\!\!\!\!\!\!
\sum_{\alpha_1\in\T{TT}_1;\;\alpha_2\in\T{TT}_n;\;\alpha_1'\in\T{TT}_1'}\!\!\!\!\!\!\!\!\!\!\!\!\!\!
\Bra{\T{S}_0}\Op{\tilde{M}}_{j_1}\Ket{\alpha_1}\nonumber\\
&\times\Bra{\alpha_1}\Op{\tilde{M}}_{j_2}\Ket{\alpha_2}\Bra{\alpha_2}\Op{\tilde{M}}_{j_3}\Ket{\alpha_1'}\Bra{\alpha_1'}\Op{\tilde{M}}_{j_4}\Ket{\T{S}_0},
\label{Eq_P_EA}
\end{align}
which are based on the Feynman diagram analysis presented in Fig.~\ref{Fig_DiagramsSpec}. Note that the strengths of both EA features are approximated by the same coefficient, Eq.~\ref{Eq_P_EA}, and that the summations explicitly include all adiabatic states underlying TT$_1$, TT$_1^*$, and TT$_n$, as well as the degenerate purely vibrational excitations represented by S$_0^*$. The peak intensities are expressed in terms of these coefficients and the triplet transition dipole moment according to
\begin{align}
&X_1=(\mu^{\T{s}_0-\T{s}_1})^4\big\vert P_\T{GB}+\big(\mu^{\T{t}_1-\T{t}_n}/\mu^{\T{s}_0-\T{s}_1}\big)^2P_\T{EA}\big\vert,\nonumber\\
&X_2=(\mu^{\T{s}_0-\T{s}_1})^4\big\vert\big(\mu^{\T{t}_1-\T{t}_n}/\mu^{\T{s}_0-\T{s}_1}\big)^2P_\T{EA}\big\vert,
\label{Eq_Peaks}
\end{align}
where the slight location mismatch between EA and GB features underlying $X_2$ is neglected. Since the coefficients can assume positive as well as negative values, taking the absolute value is necessary to yield the amplitude. Based on the measurements, the amplitude of $X_2$ amounts to roughly one third of the $X_1$ amplitude, and in our numerical evaluation, $\mu^{\T{t}_1-\T{t}_n}$ is consistently adjusted such that this ratio is maintained throughout. Fig.~\ref{Fig_FTMapAnalysis} shows a bar graph containing the calculated ratios between $P_\T{EA}$ and $P_\T{GB}$, and the values of $\mu^{\T{t}_1-\T{t}_n}$ obtained through this procedure. It is important to note that the triplet transition dipole moment obtained upon keeping the model unaltered is $\mu^{\T{t}_1-\T{t}_n}\approx15\mu^{\T{s}_0-\T{s}_1}$, close to the value applied in the foregoing calculations. This indicates that the approximate method employed here yields results consistent with explicit calculations of the FT amplitude maps.

The first two material properties that expectedly affect the peak intensities through their dependence on EA signals are the HR factors of t$_1$ and t$_n$, since the triplet transition is weighed by the vibrational overlaps between initial and final states. The t$_1$ HR factor was taken to be equal to the s$_1$ HR factor, $\lambda_{\T{t}_1}^2=\lambda_{\T{s}_1}^2=1.1$. As discussed in I, the rationale behind this is that calculations in tetracene have yielded a t$_1$ HR factor comparable to that of s$_1$,\cite{Ito_15a} and that tetracene and pentacene are expected to behave in a similar fashion in this respect. Nevertheless, given these considerations, the value $\lambda_{\T{t}_1}^2=1.1$ is somewhat arbitrary and it is therefore worthwhile to explore alternative HR factors of t$_1$, under the constraint that values do not strongly deviate from $\lambda_{\T{s}_1}^2$. Upon doing so, we confirm the dependence of the $X_1$ and $X_2$ peak intensities on this parameter, and find a significant enhancement in $P_\T{EA}/P_\T{GB}$ for $\lambda_{\T{t}_1}^2=0.8$.

For the HR factor of t$_n$ we have used an \textit{ad hoc} value of $\lambda_{\T{t}_n}^2=0$ in the absence of exemplary values reported in the literature. Interestingly, upon varying $\lambda_{\T{t}_n}^2$ (while leaving $\lambda_{\T{t}_1}^2=1.1$ unaltered) we find a region of parameter space where the vibrational overlaps between t$_1$ and t$_n$ invert the sign of $P_\T{EA}$. Consequently, the EA and GB coefficients contributing to peak $X_1$ interfere destructively, see Eq.~\ref{Eq_Peaks}. As a result, the $X_1$ to $X_2$ peak intensity ratio can be reproduced using significantly smaller values of $\mu^{\T{t}_1-\T{t}_n}$. This effect is nicely illustrated in Fig.~\ref{Fig_FTMapAnalysis}, where the case is shown for which $\lambda_{\T{t}_n}^2$ is set to 2.9, the value that minimizes the triplet transition dipole moment (referred to as ``optimal''). Although the enhancement in $P_\T{EA}/P_\T{GB}$ is modest, its sign change induces a sizable decrease in $\mu^{\T{t}_1-\T{t}_n}$.

The third material property that may contribute to the $X_1$ and $X_2$ peak intensities is dipole-dipole interaction between the two higher-lying triplet pair states that couple optically to the same pair state in the singly-excited manifolds. More specifically, when considered in the purely electronic representation, the two states $\Ket{\BS{\T{t}_1}_m,\BS{\T{t}_n}_{m'}}$ and $\Ket{\BS{\T{t}_n}_m,\BS{\T{t}_1}_{m'}}$ both couple to $\Ket{\BS{\T{t}_1}_m,\BS{\T{t}_1}_{m'}}$ through $\mu^{\T{t}_1-\T{t}_n}$, as a result of which excitation energy can be radiatively transferred between both doubly-excited states. This is very similar to the way ``conventional'' dipole-dipole interactions transfer singlet excitations through their optical coupling to the ground state. Within the commonly practiced point-dipole approximation,\cite{Fidder_91a} the associated interaction strength scales as the product of the involved transition dipole moments, $J_{m,m'}\propto\mu_m\mu_{m'}$. Hence, provided that $\mu^{\T{t}_1-\T{t}_n}$ is substantially larger than $\mu^{\T{s}_0-\T{s}_1}$, dipole-dipole interactions among triplet excitations are expectedly much stronger than those among singlet excitations. The result is a delocalization of the doubly-excited state over two diabatic pair states $\Ket{\BS{\T{t}_1}_m,\BS{\T{t}_n}_{m'}}$ and $\Ket{\BS{\T{t}_n}_m,\BS{\T{t}_1}_{m'}}$, which, in combination with vibronic coupling, leads to a redistribution of oscillator strengths of the triplet transitions, akin to vibronically coupled H- and J-aggregates.\cite{Spano_10a}

In order to explore the effect of triplet dipole-dipole couplings, we add the doubly-excited Hamiltonian with the term
\begin{align}
H_{J^\T{t}}=J^\T{t}_{m,m'}\Ket{\BS{\T{t}_1}_m,\BS{\T{t}_n}_{m'}}\Bra{\BS{\T{t}_n}_m,\BS{\T{t}_1}_{m'}},
\end{align}
where $J^\T{t}_{m,m'}$ denotes the triplet dipole-dipole interaction between molecular sites $m$ and $m'$. In order to motivate the coupling values, we first recall from I that the dipole-dipole interaction strength among nearest-neighbor singlets in pentacene crystals is $\sim8$~meV (including the effect of dielectric screening). Considering the square dependence of such couplings on the involved transition dipole moments, and the (relative) values quoted for $\mu^{\T{t}_1-\T{t}_n}$ in the literature,\cite{Pabst_08a, Zimmerman_11a, Anger_12a, Bakulin_16a} we apply a conservative value of -125~meV for nearest-neighbor couplings in the $(a,b)=(1/2,1/2)$ and $(-1/2,1/2)$ directions (other couplings need not be defined due to the spatial truncation of the triplet pairs). The minus sign is a consequence of the parallel polarization of the triplet transition dipole moments, for which the point-dipole approximation predicts couplings to be negative.\cite{Fidder_91a} As shown in Fig.~\ref{Fig_FTMapAnalysis}, incorporating these triplet dipole-dipole interactions enhances $P_\T{EA}/P_\T{GB}$ by 50~\%, with a concomitant decrease of the predicted value of $\mu^{\T{t}_1-\T{t}_n}$.

Finally, we have calculated the EA and GB coefficients while implementing all three modifications discussed above, which yields a triplet transition dipole moment close to 8. Hence, by varying material properties that have remained largely undetermined, we find about a 50~\% decrease in $\mu^{\T{t}_1-\T{t}_n}$ upon retaining an agreement for the FT amplitude maps. Conversely, for a given $\mu^{\T{t}_1-\T{t}_n}$, these material properties act alongside the triplet transition dipole moment in determining the strength of EA contributions relative to those from GB. To further substantiate the significance of these effects, further research is needed in order to pinpoint the triplet HR factors and magnitude of triplet dipole-dipole couplings. We note, however, that the material properties apart from $\lambda_{\T{t}_1}^2$ pertain to the doubly-excited manifold, and as such are not involved in the actual fission dynamics. Instead, their main relevance is centered around the spectroscopic observation of singlet fission.

\section*{Conclusions}\label{Sec_Conclusions}

The first article in this series formulated a vibronic exciton theory of singlet fission that encompasses a microscopic set of interacting electronic and vibrational degrees of freedom, treated at a non-perturbative level. In the present article, we outlined an extension of this theory which enables us to simulate 2DES of fission materials. As such, this work follows up on the recently reported first application of this optical technique to study singlet fission, and in particular on the direct detection of the optical transition between the ground state S$_0$ and the correlated triplet pair state TT$_1$ in polycrystalline pentacene. Firstly, our simulation scheme allowed us to validate our model through a comparison of simulated 2D spectra to these measurements. The overall agreement was found to be quite good, providing a firm base for the applicability of our theory.

Secondly, through a detailed analysis of our spectral data, we obtained information about TT$_1$ and the associated photophysics that is not directly accessible through experiments. Special attention was paid to the optical transitions between TT$_1$ and higher-lying triplet states TT$_n$, through which the correlated triplet pair state is commonly observed in time-resolved spectroscopy. We addressed a controversy in the literature regarding the nature of the involved higher-lying triplet states, for which different identifications, transition energies, and transition dipole moments have been reported. Within the spectral range of the ground state transitions towards TT$_1$ and S$_1$, we found that all experimental features are accounted for by including in our model only a single TT$_n$ state, using a transition energy consistent with high-level calculations for $n=2$.\cite{Pabst_08a, Khan_17a} Based on agreement with 2DES, we find that the associated transition dipole moment, $\mu^{\T{t}_1-\T{t}_n}$, is 8 to 15 times larger than the singlet transition dipole moment $\mu^{\T{s}_0-\T{s}_1}$. We nevertheless emphasize that our determination of $\mu^{\T{t}_1-\T{t}_n}$ is qualitative rather than quantitative. In addition, we have brought forward several material properties that impact the optical transition strength between TT$_1$ and TT$_n$ in addition to $\mu^{\T{t}_1-\T{t}_n}$, and which consequently interfere with a direct determination of this parameter based on spectroscopy. Among the contributing factors, we have found the HR factors of t$_1$ and t$_n$ to be of importance, as is the dipole-dipole interaction between higher-lying triplets.

Most of the aspects discussed here pertain to the optical transition between TT$_1$ and TT$_n$, which is of critical importance to the detection of singlet fission through time-resolved spectroscopy. Nevertheless, these aspects do not impact the process of fission itself. Having passed the bar set by successfully comparing simulated 2D spectra to experimental results, we will return to this process in a forthcoming article, where dynamical simulations of fission materials will be presented. In doing so, we will focus in particular on the role of high-frequency vibrational modes in facilitating ultra-fast exciton multiplication in crystalline pentacene, with the aim of unraveling principles that are common to fission materials in general.

\section*{Acknowledgements}

The authors wish to thank Artem Bakulin, Akshay Rao, and Donatas Zigmantas for sharing data of the 2DES measurements on polycrystalline pentacene. R.T.~acknowledges The Netherlands Organisation for Scientific Research NWO for support through a Rubicon grant. D.R.R.~acknowledges funding from NSF grant no.~CHE--1464802.


%

\end{document}